\documentclass[12pt,preprint]{aastex}
\shorttitle{The Young Cluster Population of M82 Region B}
\shortauthors{Smith et al.}
\begin{document}
\title{The Young Cluster Population of M82 Region B\altaffilmark{1}} 
\author{L.J. Smith\altaffilmark{2,3}}
\author{N. Bastian\altaffilmark{3}}
\author{I.S. Konstantopoulos\altaffilmark{3}}
\author{J.S. Gallagher III\altaffilmark{4}}
\author{M. Gieles\altaffilmark{5}}
\author{R. de Grijs\altaffilmark{6,7}}
\author{S.S. Larsen\altaffilmark{8}}
\author{R.W. O'Connell\altaffilmark{9}}
\author{M.S. Westmoquette\altaffilmark{3}}
\altaffiltext{1}{Based on observations with the NASA/ESA {\it Hubble
    Space Telescope}, obtained at the Space Telescope Science
    Institute, which is operated by AURA, Inc., under NASA contract
    NAS5--26555. These observations are associated with program \#10853.} 
    \altaffiltext{2}{Space Telescope Science Institute and European Space Agency, 3700 San Martin Drive,
  Baltimore, MD 21218; lsmith@stsci.edu}
\altaffiltext{3}
{Department of Physics and Astronomy, University College London, Gower St., London WC1E 6BT, UK}
\altaffiltext{4}{Department of Astronomy, University of Wisconsin-Madison, SAL Chamberlin Halll,
1150 University Ave, Madison WI 53706}
\altaffiltext{5}{European Southern Observatory, Casilla 19001, Santiago 19, Chile}
\altaffiltext{6}{Department of  Physics and Astronomy, The University of Sheffield,
Hicks Building, Hounsfield Rd., Sheffield, S3 7RH, UK}
\altaffiltext{7}{National Astronomical Observatories, Chinese Academy of Sciences, 20A Datun Road, Chaoyang District, Beijing 100012, P.R. China}
\altaffiltext{8}{Astronomical Institute, Utrecht University, Princetonplein 5, NL-3584 CC Utrecht, The Netherlands}
\altaffiltext{9}{Department of Astronomy, University of Virginia, 
P.O. Box 3818, Charlottesville, VA 22903}
\begin{abstract}
We present observations obtained with the Advanced Camera for Surveys on board the Hubble Space Telescope of the ``fossil'' starburst region B in the nearby starburst galaxy M82. By comparing $UBVI$ photometry with models, we derive ages and extinctions for 35 $U$-band selected star clusters. We find that the peak epoch of cluster formation occurred $\sim 150$ Myr ago, in contrast to earlier work that found a peak formation age of 1.1 Gyr.  The difference is most likely due to our inclusion of $U$-band data, which are essential for accurate age determinations of young cluster populations. We further show that the previously reported turnover in the cluster luminosity function is probably due to the neglect of the effect of extended sources on the detection limit. The much younger cluster ages we derive clarifies the evolution of the M82 starburst. The M82-B age distribution now overlaps with the ages of: the nuclear starburst; clusters formed on the opposite side of the disk; and the last encounter with M81, some 220 Myr ago.
\end{abstract}
\section{Introduction} \label{intro}

The star cluster population in Region B of the starburst galaxy M82 has received a large amount of attention in recent years because of claims that it is an intermediate-aged ($\sim 1$ Gyr) system with a cluster luminosity function (CLF) that is fundamentally different to those found in galaxies containing younger star clusters (de Grijs, Bastian \& Lamers 2003b; hereafter GBL03b). Region B in M82 was first cataloged by O'Connell \& Mangano (1978) as the visually brightest region outside of the nucleus; it lies 0.5--1 kpc north-east of the galaxy center. They found that the intrinsic brightness of this region at an age of 10 Myr rivaled that of the present-day nuclear starburst. The formation of the younger stellar populations in region B was therefore an exceptional event in the star formation history of M82 and is presumably linked to a past close encounter with its neighbor M81 (e.g. Brouillet et al. 1991; Yun, Ho \& Lo 1994).

The cluster population of M82-B was first studied by de Grijs, O'Connell \& Gallagher (2001; hereafter GOG01) using $BVI$  and $JH$ images obtained with the Wide Field and Planetary Camera 2 (WFPC2) and the Near-Infrared Camera and Multi-Object Spectrometer (NICMOS) on board the Hubble Space Telescope (HST). de Grijs, Bastian \& Lamers (2003a; hereafter GBL03a) and GBL03b derived ages and extinctions for the cluster sample using $BVIJH$ photometry and found a peak cluster formation epoch of 1.10 Gyr with 42 out of 80 clusters formed between 500 Myr and 1.5 Gyr. They correct the luminosities of this subset to a fiducial age of 1.0 Gyr and find that the resulting CLF has a log-normal shape. Their interpretation is that the M82-B CLF has the same characteristic shape as that found for globular clusters (GCs) rather than the power-law CLF found for young massive cluster (YMC) populations (Elson \& Fall 1985; Elmegreen \& Efremov 1997; Whitmore et al. 2002; Bik et al. 2003; Larsen 2007). Critically, if YMCs are to evolve eventually into GCs, the CLF must also change with time. Thus, the shape of the CLF for the rare intermediate age M82-B cluster population is crucial for theoretical studies of the evolution of the CLF. 

Although much research effort has been focussed on the evolution of the CLF, the results are not yet conclusive. Most
attempts to understand the mass functions of old GCs start from the assumption that any initial cluster mass function (CMF which is related to the CLF) will be eventually transformed into a Gaussian distribution because of the preferential disruption of the low mass clusters through, for example, tidal interactions with the host galaxy (e.g. Elmegreen \& Efremov 1997; Gnedin \& Ostriker 1997; Fall \& Zhang 2001; Vesperini \& Zepf 2003). de Grijs, Parmentier \& Lamers (2005a), however, find that the initial mass distribution of the M82-B clusters, based on their previous data and the assumption that the cluster disruption time is set by the galaxy environment, is not consistent with an initial power law distribution.

In this paper, we re-examine the ages and, most importantly, extinctions of the M82-B cluster population using new observations obtained with the Advanced Camera for Surveys (ACS) on board HST. 
\section{Observations} \label{obs}

Observations with the High Resolution Channel (HRC) of the ACS on board HST were made of region B on UT 2006 November 4 and 5 using the F330W($U$) filter (Prop ID: 10853, PI L.J. Smith). These observations were designed to be supplementary to the M82 mosaic obtained by the Hubble Heritage Team using the Wide Field Channel (WFC) of the ACS and the F435W ($B$), F555W($V$) and F814W($I$) filters (Mutchler et al. 2007). As shown by Anders et al. (2004), four is the minimum number of passbands needed to determine reliable photometric ages and extinctions from optical spectral energy distributions (SEDs), with $U$ and $B$ being crucial for ages $\le$ few Gyr. Note that we adopt the Johnson $UBVI$ notation for simplicity but we do not convert any ACS magnitudes derived in this paper to a standard ground-based $UBVI$ system.

We followed the proven observational methodology of GOG01 and imaged two adjacent $29'' \times 26''$ fields to cover the full extent of region B. The two sub-regions ``B1'' and ``B2'' have field centers located at RA$=09^{\rm h}\,56^{\rm m}\,01^{\rm s}.5$; Dec$=+69^\circ\,41'\,09''.7$  and  RA$=09^{\rm h}\,55^{\rm m}\,56^{\rm s}.2$; Dec$=+69^\circ\,41'\,01''.9$ (J2000). The total exposure time was 8331 s for each sub-region and we used a dither pattern to optimize removal of hot pixels, cosmic rays and HRC artifacts.  The data were retrieved and automatically calibrated on-the-fly from the Multimission Archive at STScI (MAST). The final images have a resolution of $0''.025$ pixel$^{-1}$ or 0.44~pc at the distance to M82 of 3.6~Mpc (Freedman et al. 1994).

\begin{figure*}
\figurenum{1}
\plotone{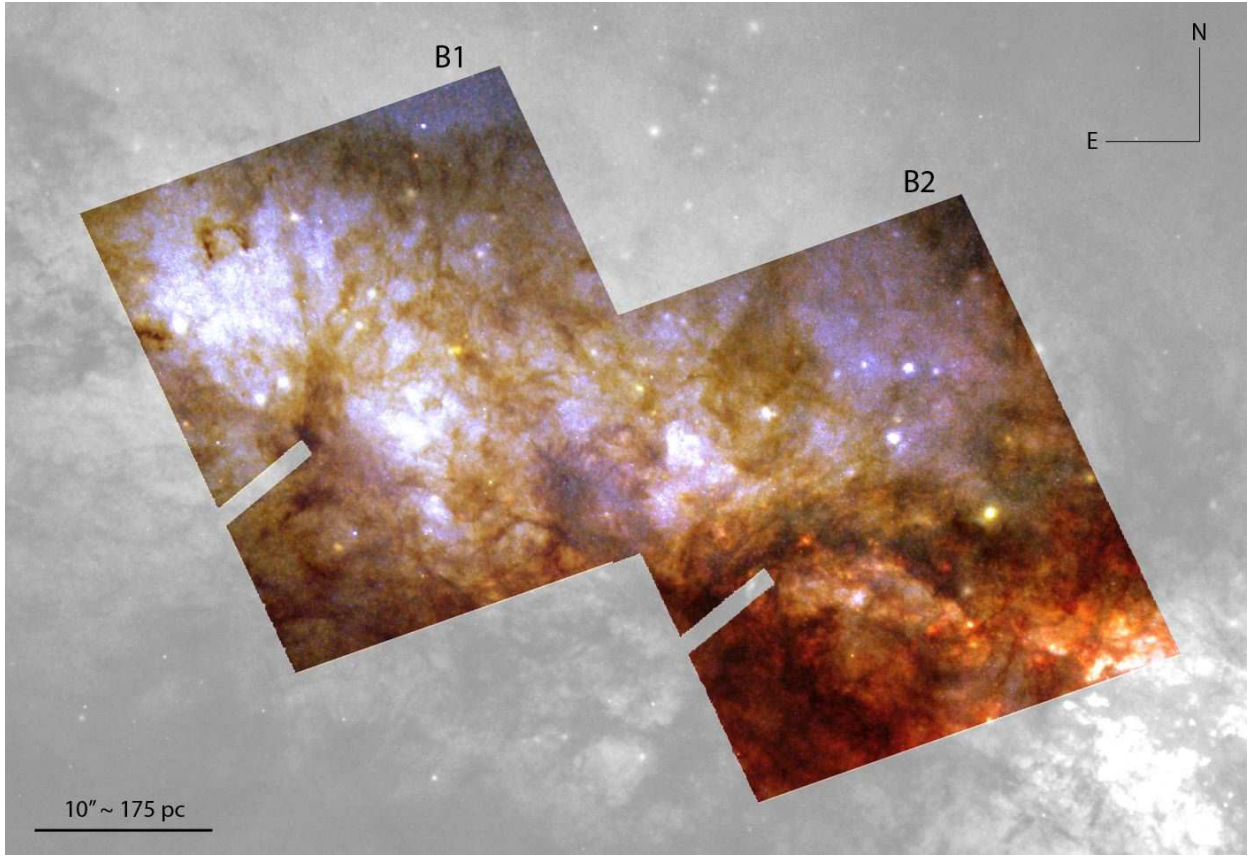}
\caption{{\sc plate 1} Composite color image of M82-B created using the HRC F330W and the WFC F555W and F814W filters, superimposed on a gray-scale F555W image of M82. Sub-region B1 is on the left and farthest from the nucleus; B2 is on the right; the edge of the nuclear starburst region A is visible in the lower right-hand corner. The finger-like shadow of the HRC occulting mask is seen extending from the left-hand edge of each image. Overall, this image illustrates the complex structure of region B and shows the highly variable extinction, and the wide range in intrinsic brightnesses of the clusters. Many of the clusters in this study are not visible here due to dynamic range limitations but they are associated with the bright blue regions which are zones of low to moderate extinction.}
\label{fig:uvi}
\end{figure*}

For the complementary ACS/WFC $BVI$ images, we used the drizzle-combined high-level science products available from MAST (pixel size$=0''.05$), as detailed by Mutchler et al. (2007). In Fig.~\ref{fig:uvi} (Plate 1), we show a color composite image of region B made with the $UVI$ filters. The bright blue regions containing most of the cluster population are surrounded by intricate dust lanes, suggesting that we are viewing the cluster population of the disk through gaps in the foreground dust layer. 
\section{Photometry, Ages and Extinctions} \label{photom}
Our sample of 35 clusters was selected by eye from the two HRC F330W images using the criteria that the objects were well-resolved and detected in the three WFC filters. After experimenting, we felt that this selection technique provided a reliable method for obtaining a sample of bona-fide clusters.
We performed aperture photometry on the HRC images using a radius of 20 pixels and a background annulus with an inner radius and width of 22 and 2 pixels respectively. 
For the WFC images, we used an aperture of the same physical size as the one used for the HRC data i.e. 10 pixels with a background annulus radius and width of 11 and 1 pixel. These values were chosen to maximize the aperture size and minimize neighboring contamination and background variations.

For both datasets, we transformed the measured flux to the {\sc vegamag} system using the zero points of Sirianni et al. (2005). Aperture corrections were derived by comparing photometry for two bright isolated clusters (\#25 and \#34) using apertures of 10 and 40 pixels for the WFC images, and 20 and 80 pixels for the HRC images. The following average aperture corrections were derived for the $UBVI$ filters: $-0.64, -0.35, -0.33$ and $-0.43$ mag. 
The photometry for the 35 clusters is presented in Table~\ref{tab:phot}. We performed completeness tests on the images by adding sources of various brightnesses and sizes and recovering them in an automated fashion.  This is discussed in more detail in Section~\ref{disc}.

GOG01 and GBL03a identify a total of 113 clusters with $V \le 22.5$ mag with 46 of these having upper limits in one or more of the Johnson $BVIJH$ filters. We find that 28 are in common with our sample. To compare our photometry with GOG01, we used the transformations given in Sirianni et al. (2005) to convert our ACS magnitudes to the Johnson $BVI$ filters. We find that in terms of the colors, the mean Johnson $(B-V)$ difference is $-0.01\pm0.10$, and for $(V-I)$ we find a mean difference of $-0.17\pm0.22$ mag with the GOG01 color being bluer.

\begin{figure}
\figurenum{2}
\plotone{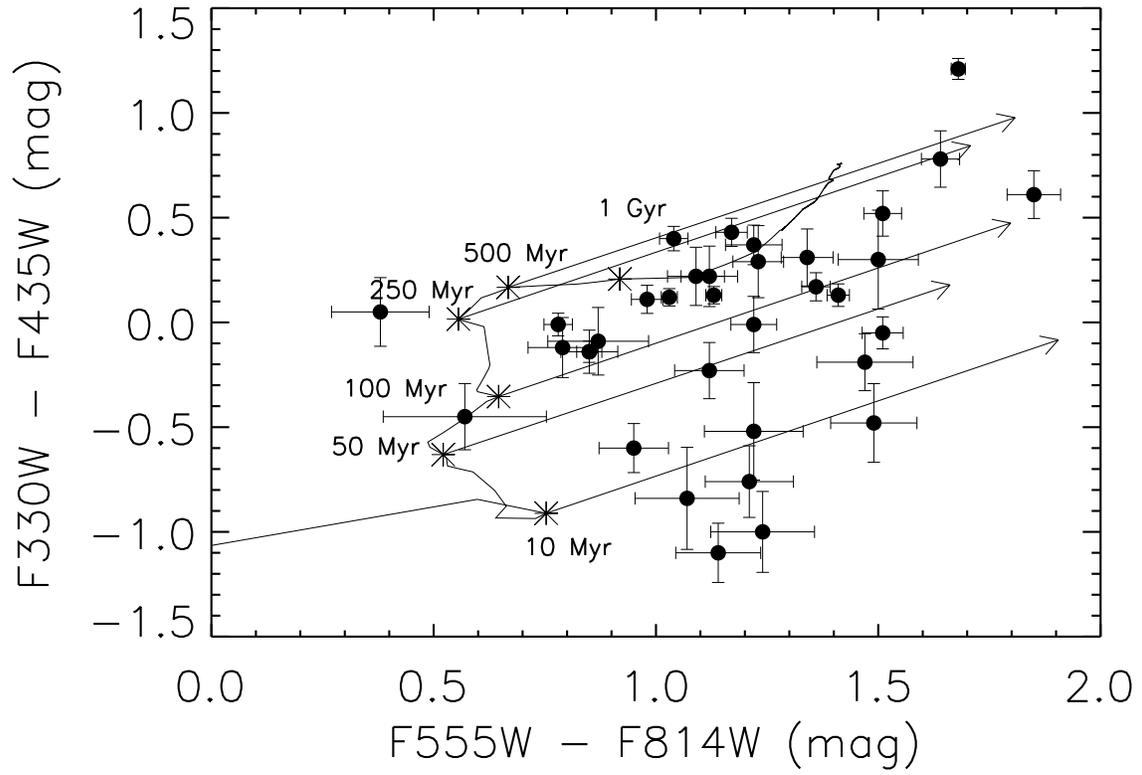}
\caption{Color-color plot showing $(F330W-F435W)$ vs. $(F555W-F814W)$ for all 35 clusters. The asterisks (joined by a solid line) indicate {\sc galev} models for ages of 10, 50, 100, 250, 500 and 1000 Myr. The arrows represent extinction vectors $A_{\rm V}$ with a length of 3 mags.
}
\label{fig:col}
\end{figure}

To determine the ages and extinctions for the sample of clusters in region B, we have compared the observed energy distributions as given by the $UBVI$ photometry to the theoretical {\sc galev} simple stellar population (SSP) models of Anders \& Fritze-v.~Alvensleben (2003). These models use stellar isochrones from the Padova group and have the advantage that they provide broad-band colors as a function of age in the {\sc vegamag} HST filter system. We adopt a Salpeter initial mass function slope, with a lower mass limit of 0.15~M$_\odot$, and solar metallicity, as this is appropriate for M82 (Origlia et al. 2004; Smith et al. 2006). In Fig.~\ref{fig:col} we show the positions of the cluster sample on a $(U-B)$ vs. $(V-I)$ color-color plot and compare them to the {\sc galev} models. The effect of extinction is also shown where we adopt the standard Galactic extinction law of Savage \& Mathis (1979). It can be seen that the clusters are reddened and most have ages of less than 1 Gyr. The one discrepant point in the upper right-hand corner is cluster \#34. A careful inspection of the HST images shows that the photometry is compromised by a dust-lane crossing the cluster (see Konstantopoulos et al. 2007 for details). We omit this cluster from further study here. 

Fig.~\ref{fig:col} illustrates only one of a series of possible projections of the photometric data in color space. The  $(U-B)$ color is particularly sensitive to clusters $<200$~Myr old whereas $(B-V)$ and $(V-I)$ colors provide a better age discrimination for older clusters. 
However, $U$-band-based colors still provide information on the extinction.
Thus to fully exploit the data and allow for the fact that SSP colors do not evolve monotonically in color-space, we have used the three-dimensional maximum likelihood method (3DEF), described in detail by Bik et al. (2003) and Bastian et al. (2005), to derive ages and extinctions for the clusters. This technique compares the photometry with the {\sc galev} models for a series of ages and extinctions,  fitting all four filters simultaneously.
The model with the lowest reduced $\chi^2_\nu$ is selected as the best fit and the errors are determined by the extrema of the models which satisfy $\chi^2_\nu < \chi^2_{\nu, {\rm best}} +1$. The derived parameters are given in Table~\ref{tab:phot}.

\begin{figure}
\figurenum{3}
\plotone{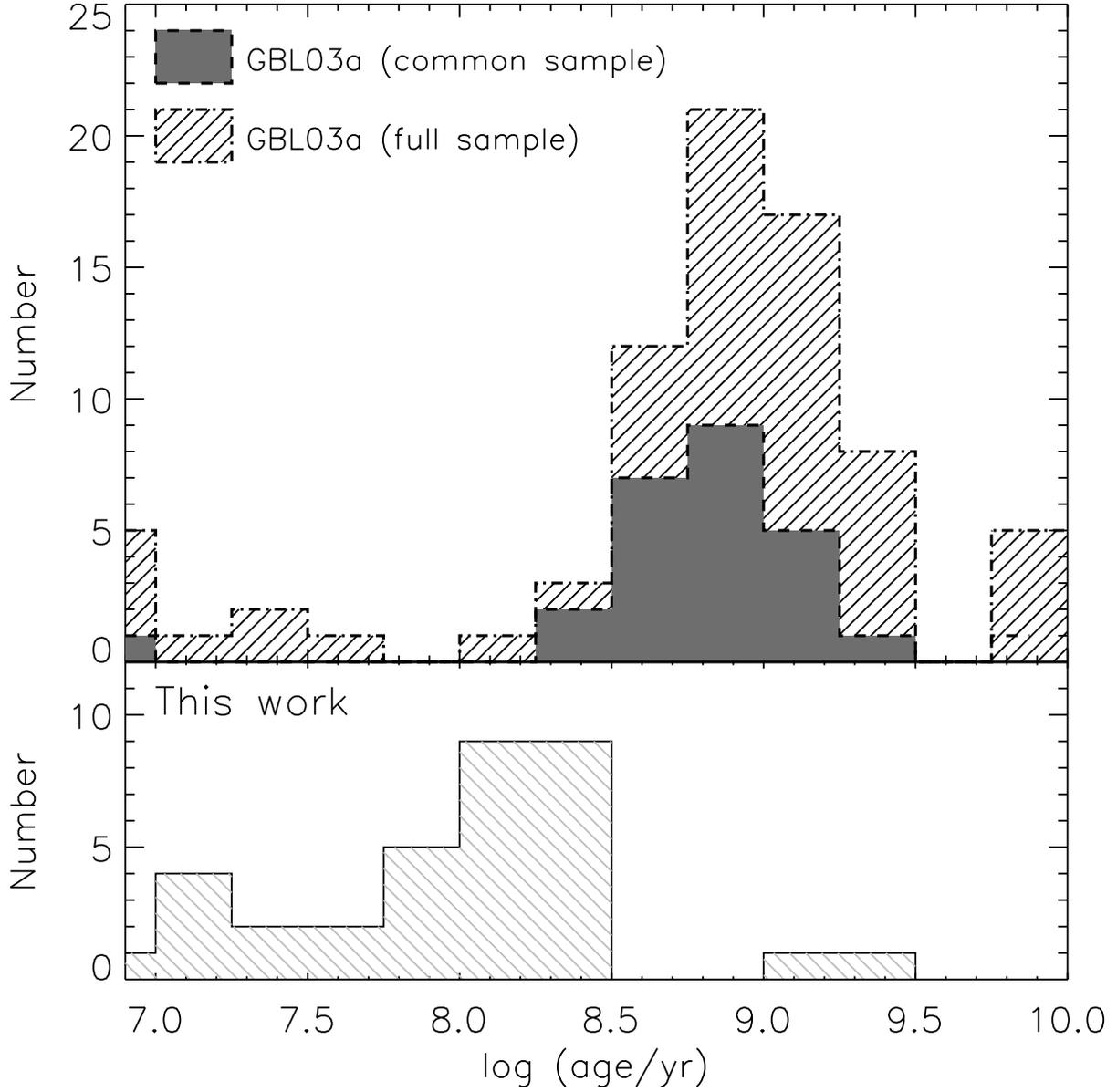}
\caption{Upper plot: The distribution of cluster ages for the sample of 26 clusters in common with GBL03a using the GBL03a ages compared with the full sample of 80 GBL03a clusters. Lower plot: The distribution of cluster ages for the sample in this paper using the ages presented in Table~\ref{tab:phot}.}\label{fig:hist}
\end{figure}

In Fig.~\ref{fig:hist} we compare the age distribution of our full sample of 35 clusters with the age distribution of GBL03a. We divide the GBL03a cluster sample into two samples: (a) the full GBL03a sample of 80 clusters that have assigned ages; and (b)
the 26 clusters from GBL03a in common with our sample of 35. It is apparent that we find considerably younger ages for our cluster sample with a peak formation epoch at log (age/yr) $=8.18$ or 150 Myr with a dispersion in age (yr) of 0.25 dex  compared to the GBL03a peak at log (age/yr) $=9.04$ or 1.10~Gyr with an age dispersion (yr) of 0.27 dex. While the median ages differ between the two studies, both studies find very young clusters with ages below 50~Myr, as shown in Fig. 3. The youngest clusters in our sample have ages of $\sim 12$~Myr, and most of these are in region B2.

Fig.~\ref{fig:hist}  also shows that the age distribution for the common GBL03a sub-sample of 26 clusters is similar to the age distribution for the full GBL03a sample. This suggests that our sample based on $U$-band selection is not heavily biased towards the youngest clusters in the GBL03a sample.
The extinctions we derive are much higher; 84\% of the GBL03a sample have $E(B-V) \le0.1$~mag with 34\% having zero reddening. In contrast we find the clusters to be fairly reddened in accord with the appearance of region B in Fig.~\ref{fig:uvi}; our mean reddening is $E(B-V)=0.39\pm0.23$~mag. 
GBL03a used the same 3DEF method employed here but SSP models based on Bruzual \& Charlot (1996). As shown by de Grijs et al. (2005b), the use of different SSP models
has very little effect on age determinations.
\section{Discussion} \label{disc}

It is well known that the effects of age and reddening are difficult to disentangle in photometric datasets with a restricted wavelength range (e.g. Anders et al. 2004). The studies of GOG01 and GBL03a,b, using $BVI$ and $BVIJH$ photometry respectively, found best fitting SED solutions with little or no reddening and ages close to 1 Gyr. With the addition of the HRC $U$-band data covering the Balmer jump region, we find solutions that favor much younger ages and higher extinctions{\footnote {If extinctions were low, we would expect to have found many more $U$-band bright clusters (Fig. 1) but these are not seen.}} for clusters in common with GBL03a. We note that our findings  cannot be explained by the Johnson $(V-I)$ color difference of $-0.17\pm0.22$ mag between the two datasets (Sect.~\ref{photom}) because the GOG01 colors are bluer and so this shift is in the wrong direction.
Optical HST spectra of the two brightest clusters in our sample (\#1 and \#34) are presented by Smith et al. (2006). They find lower ages and higher extinctions compared to GOG01 and GBL03a.
Konstantopoulos et al. (2007) have obtained Gemini spectroscopy for six clusters in common with our sample; comparison of the spectroscopically and photometrically derived ages shows very good agreement. We are thus confident that our ages and reddenings are robust. 
It is not clear why the studies of GOG01 and GBL03a found lower extinctions. One possibility is that the M82 extinction law differs from the Galactic one at longer wavelengths. Another possibility is that SSP models are less accurate in the near-IR.

We now consider whether the M82-B cluster luminosity function has a log-normal distribution, as claimed by GBL03b. In this work, the CLF was determined using 42 clusters with well-determined ages corrected to a common age of 1.0 Gyr. A clear turnover is observed 2 mag above their 100 per cent detection limit at $V=22.5$ mag for a point source. de Grijs et al. (2005a) consider the equivalent detection limit for an extended source with a typical cluster size of 5~pc, and find that it has little effect ($\lesssim 0.5$~mag) on the turnover magnitude. They also consider whether an artificial turnover could be caused by variable extinction across region B, and find that they can rule this out unless the extinction for most clusters is $A_{\rm V} > 1$ mag, as we find here.

\begin{figure}
\figurenum{4}
\plotone{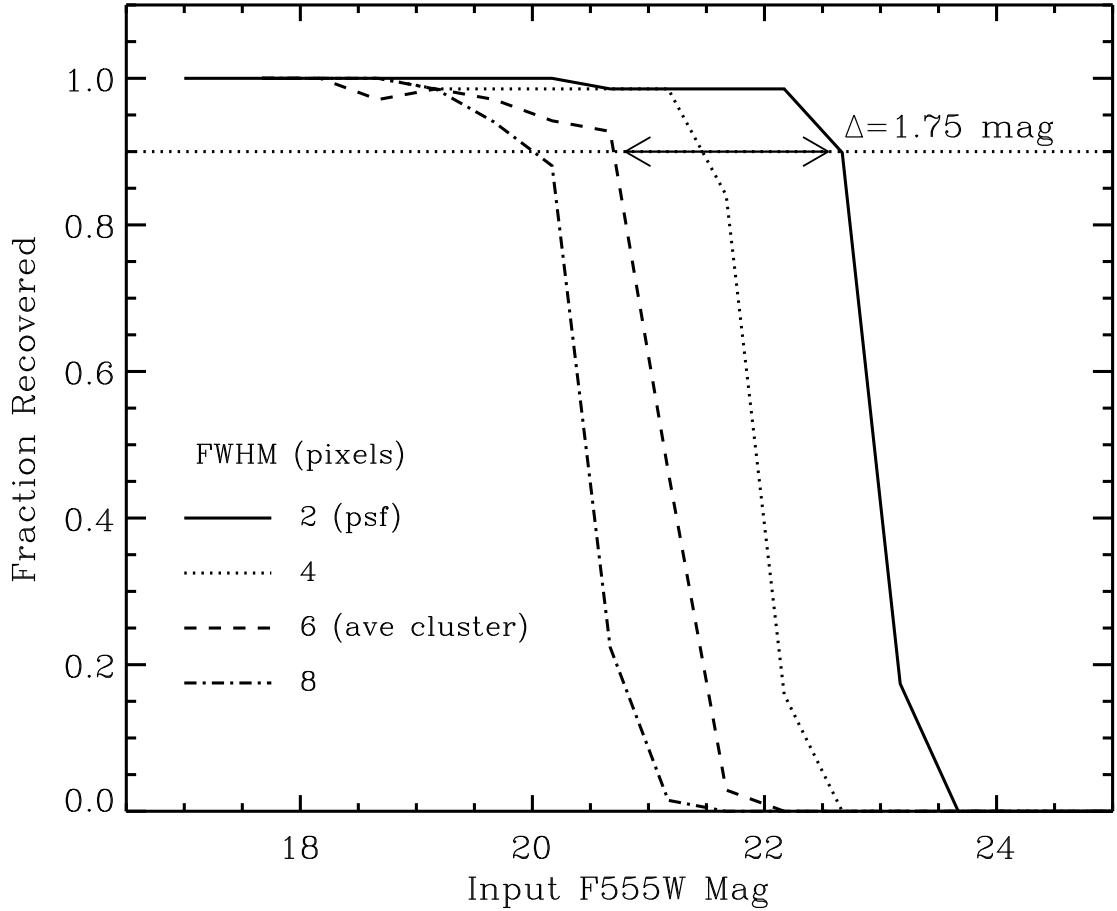}
\caption{Recovered fraction of artificial sources as a function of input magnitude where the recovered fraction is normalised to the number of clusters recovered at the brightest magnitude used. The 90\%\ detection limit will shift by 1.75 mag if the sources have a typical cluster size of 6 pixels at the distance of M82.}
\label{fig:det}
\end{figure}

We first examine the non-trivial effect of an extended source compared to a point source on the detection limit. We added 150 artificial sources to the F555W image within region B, with all clusters being given the same magnitude and size per trial.  The clusters were then recovered using SExtractor (Bertin \& Arnouts 1996), and this process was carried out for various magnitudes and extinctions. In Fig.~\ref {fig:det}, we show the recovered fraction of artificial sources as a function of input magnitude for point sources (FWHM$=2$ pixels) and extended sources with FWHMs up to 8 pixels (where 1 PC or WFC pixel $=0''.05 = 0.9$ pc). It can be seen that the detection limit becomes 1.75~mag brighter for a cluster with a typical FWHM of 6 pixels or 5~pc at the distance of M82 compared to an equivalent point source. A similar conclusion on completeness limits for extended objects is reached by Bastian et al. (2005) and Mora, Larsen \& Kissler-Patig (2007). We thus conclude that the turnover in the M82-B CLF as presented by GBL03b is most likely due to the detection limit of the cluster sample. The much larger and variable extinction that we have found for the M82-B clusters will also affect the CLF, as discussed by GBL03b.

 We note that the derivation of the true CLF for the M82-B clusters is difficult for at least two reasons. First, each cluster will have its own detection limit in each filter depending on its size. This effect is significant because of the proximity of M82. Second, the clusters we detect in region B are simply the least obscured ones, thus we may not be seeing a representative population.

Finally, we compare our peak epoch of cluster formation at 150~Myr with the date of the last encounter of M82 with M81. The most recent and detailed simulations are presented by Yun (1999) who used an updated version of the Brouillet et al. (1991) code
which includes tidal interactions with a third galaxy NGC 3077. The derived time scale for the last encounter of M82 with M81 is given as 220 Myr, suggesting that the M82-B cluster population formed as a direct result of this interaction (see also GOG01).
\section{Conclusions}
We have presented new HST/ACS $UBVI$ photometry for 35 $U$-band selected massive star clusters in the post-starburst region B of M82. We find that our sample of clusters are more reddened and have significantly younger ages than the previous study of GBL03a which did not have access to $U$-band data. We determine that the peak epoch of cluster formation for this sample occurred  close to 150~Myr ago. Consistent with GOG01, we find that star formation continued in M82-B until $\sim$ 12--20~Myr ago. We suggest that the turnover in the CLF found by GBL03b is an artifact caused by underestimating the detection limit for well-resolved clusters.  The M82-B clusters were probably formed as a result of the last encounter with M81 $\sim 220$ Myr ago (Yun 1999). The much younger ages that we find makes the M82 starburst easier to understand as an evolving system. 

The M82-B cluster age distribution overlaps with the 8--15~Myr old nuclear starburst (F\"orster Schreiber et al. 2003), and with the 60~Myr old clusters M82-F and L located on the opposite side of the disk to M82-B (Gallagher \& Smith 1999). Although the disk and nuclear starburst zone clearly have different histories, this age overlap removes the distinction that M82-B is a special, offset, fossil starburst region within M82. Instead we suggest that M82-B provides a uniquely low obscuration view of a typical region of the inner M82 stellar disk, allowing us to glimpse the population of star clusters produced following its last interaction with M81. This model is further supported by Konstantopoulos et al. (2007) where we present spectroscopy and radial velocities for seven M82-B clusters, and find agreement between the spectroscopic and photometric cluster ages.
\acknowledgments
Support for program \#10853 was provided by NASA through a grant from the Space Telescope Science Institute, which is operated by AURA, Inc., under NASA contract NAS 5-26555.


\clearpage

\begin{deluxetable}{lccccccccccc}
\tabletypesize{\scriptsize}
\tablenum{1}
\tablecaption{Photometry and Derived Parameters of the Cluster Sample in Region B
\label{tab:phot}}
\tablehead{
\colhead{ID}&
\colhead{F330W} &
\colhead{F435W} &
\colhead{F555W} &
\colhead{F814W} &
\colhead{$\chi^2_\nu$} & 
\multicolumn{3}{c}{$E$(F435W$-$F555W)} &
\multicolumn{3}{c}{Log Age}
\\
\colhead{}&
\colhead{(mag)}&
\colhead{(mag)}&
\colhead{(mag)}&
\colhead{(mag)}&
\colhead{}&
\multicolumn{3}{c}{(mag)} &
\multicolumn{3}{c}{(yr)} \\
&&&&&&
\colhead{Min}&
\colhead{Best}&
\colhead{Max}&
\colhead{Min}&
\colhead{Best}&
\colhead{Max}
\\ [-2.5ex]
}
\startdata
   1 &18.37 $\pm$ 0.01 &18.24 $\pm$ 0.04 &17.74 $\pm$ 0.03 &16.61 $\pm$ 0.02
    &  0.6 &  0.34 &  0.38 &  0.46 &  8.03 &  8.22 &  8.28
    \cr
   2 &19.64 $\pm$ 0.04 &20.24 $\pm$ 0.11 &19.85 $\pm$ 0.10 &18.90 $\pm$ 0.12
  &  0.1 &  0.20 &  0.28 &  0.40 &  6.90 &  7.45 &  7.75 
    \cr
  3 &20.30 $\pm$ 0.10 &20.82 $\pm$ 0.21 &20.40 $\pm$ 0.18 &19.18 $\pm$ 0.13
   & 0.1 &  0.30 &  0.42 &  0.54 &  6.90 &  7.38  &  7.78 
   \cr
  4 &20.03 $\pm$ 0.06 &20.15 $\pm$ 0.13 &19.75 $\pm$ 0.11 &18.96 $\pm$ 0.11
  &  0.1 &  0.06 &  0.16 &  0.38 &  7.78 &  8.23 &  8.46 
   \cr
    5 &19.42 $\pm$ 0.03 &19.02 $\pm$ 0.05 &18.49 $\pm$ 0.05 &17.45 $\pm$ 0.04
    &  0.4 &  0.14 &  0.34 &  0.42 &  8.29 &  8.49 &  8.89 
     \cr
\enddata
\tablecomments{[The complete version of this table is in the electronic edition of the Journal. The printed edition contains only a sample.]}
\end{deluxetable}
%
\clearpage
\end{document}